\documentclass[12pt,a4paper]{article}
\usepackage{amssymb}
\usepackage[T2A]{fontenc}
\usepackage[cp866]{inputenc}
\usepackage[russian,english]{babel}
\usepackage[unicode]{hyperref}
\begin{document}
\def\emline#1#2#3#4#5#6{%
       \put(#1,#2){\special{em:moveto}}%
       \put(#4,#5){\special{em:lineto}}}
\def\newpic#1{}
\voffset=-2cm
\begin{center}
{\Large  
The pedagogical value of the four-dimensional picture II: Another way of looking 
at the electromagnetic field}\\[3mm]
{B. P. Kosyakov} \\[3mm] 
{ Russian Federal Nuclear Center--VNIIEF, and Sarov Institute of Physics {\&} Technology, Sarov, 607190 Nizhnii Novgorod 
Region, Russia\\
} 
\end{center}
\begin{abstract}
\noindent
{\small 
The concept of electromagnetic field can be neatly formulated by recognizing that the simplest  
form of the four-force is indeed feasible. 
We  show that Maxwell's equations almost entirely stem from the properties of spacetime, notably from the fact that our world has dimension $d=4$.
Their complete reconstruction requires three additional assumptions which are seemingly divorced from geometry, but, actually, may have much to do with the spacetime properties.}
\end{abstract}

\section{Introduction}
\label
{Introduction}
In this, second, paper of a series of papers, initiated by Ref.~\cite{k2013}, we review the utility of some basic four-dimensional concepts 
in classical electrodynamics. 

In Section 2, we discuss the question of {whether the idea of electromagnetic field} can be made a neat theoretical concept.
The student embarking on a study of classical electrodynamics in the course of theoretical physics may be surprised that {textbooks do not begin with a definition of the electromagnetic field}.
Strange as it may seem, this definition is also missed from the later stages of the physics curriculum.
For example, a graduate course in electromagnetism covered by the Jackson's book \cite{Jackson}, which is informally accepted as {\it the} standard textbook of classical electrodynamics in western universities, is lacking in a well defined concept of electromagnetic field.
To introduce the electric and magnetic field strengths \textbf{E} and \textbf{B}, Jackson turns to the Lorentz force equation\footnote{In this paper, we use the Gaussian system of units, and put the speed of light to be 1.} 
\begin{equation}
{\bf F}=q\left({\bf E}+{\bf v}\times{\bf B}\right),                
\label
{lorentz tree-force}
\end{equation}                       
and notes that this equation was originally discovered experimentally and then justified in numerous experimental studies \cite{Jackson3}.
The impression gained from such an introduction of \textbf{E} and \textbf{B} is that the defining equation (\ref{lorentz tree-force}) is well substantiated phenomenologically but has no 
{theoretical} motivation.
Why is the {theoretical} motivation necessary? 
A phenomenological equation may be found to be merely an approximation for a fundamental relationship, and if one day we would have to correct the analytic form of equation (\ref{lorentz tree-force}), then the very sense of what we held to be the electromagnetic field may drastically change along with it.
A strongly motivated and uniquely formulated definition is necessary because otherwise we discuss the properties of a physical object being in the dark about its essence, and hence run the risk to confuse it with objects of another nature.

The desired definition of the electromagnetic field can be found by recognizing that the {\it simplest} analytic form of the four-force $f^\mu$ is indeed feasible.
Furthermore, we will see that the associated three-force \textbf{F} eludes the simplicity argument, which offers an explanation for why equation  (\ref{lorentz tree-force}) fails to be the defining equation of the electromagnetic field.

Section 3 is devoted to the dynamical law for the electromagnetic field.
We show that Maxwell's equations are almost entirely derivable from the 
properties of spacetime.
The complete reconstruction of Maxwell's equations requires three additional assumptions 
which are seemingly divorced from geometry.
However, a closer inspection of symmetries peculiar to electrodynamics in the spacetime
dimension $d=4$ makes it clear that these allegedly non-geometric assumptions may have 
much to do with geometry.

There are many suggestions as to how Maxwell's equations can be derived with the use of plausible,  easy, and elegant heuristic considerations (see, e.~g., 
Ref.~\cite{Diener} and references therein).
However, the plausibility, easiness, and  elegance by themselves are of little concern in the present approach.
An effort is made to understand to {what extent the form of the dynamical law governing the electromagnetic field is ordered by geometrical features of our world}, in particular by the fact that {spacetime has four dimensions}.

\section{What is the electromagnetic field? }
\label
{definition}

Let us clarify what is meant by the {definition} of electromagnetic field.
If the concept of a particle is primary, then the  
definition should display how the field exerts on particles.
The key observation is the fact that the four-force  $f^\mu$ is orthogonal to the four-velocity 
of a particle $v^\mu$ at the point where this four-force is applied, and hence $f^\mu$ cannot be independent of $v^\mu$. 
It is then reasonable to begin with looking for the {\it simplest} form of this dependence.  
Let $f^\mu$ be  {\it linear} in $v^\mu$. 
The case $f^\mu=\alpha v^\mu$ is of no interest because   
$f^\mu$ is orthogonal to $v^\mu$ only for $\alpha=0$.  
Hence, we turn to the construction
\begin{equation}
f^\mu=\beta^{\mu\nu}v_\nu\,\,.
\label
{f mu=beta mu nu v_nu}
\end{equation}
By the condition of orthogonality  
\begin{equation}
f\cdot v=0\,\,,
\label
{fv=0}
\end{equation}
we come to the equation
\begin{equation}
\beta^{\mu\nu}v_\mu v_\nu=0\,\,,
\label
{beta mu nu v_nu v_mu}
\end{equation}
which is obeyed by an arbitrary $v^\mu$ provided that $\beta^{\mu\nu}=-\beta^{\nu\mu}$.
Therefore, if there exists a physical object which is distributed over all space and affects any particle through a four-force linear in the four-velocity, then this object is characterized by an {antisymmetric} tensor $\beta^{\mu\nu}$ at each point.
Such objects are collectively known as {force fields}, or simply {fields}.
Note, however, that $\beta^{\mu\nu}$ contains information on both the state of the field and how it affects the particle. 
Let us separate these concepts.
Consider the {\it simplest case} that the coupling of a particle with the field is given by
a  real {\it scalar} parameter $q$.
Then (\ref{f mu=beta mu nu v_nu}) becomes 
\begin{equation}
{f^\mu}=qv_\nu F^{\mu\nu}\,\,.
\label
{f mu=eF mu nu v_nu}
\end{equation}
This possibility is actually realized in nature.
The field whose state at each point $x^\mu$ of Minkowski space is specified by an {\it antisymmetric} tensor $F_{\mu\nu}$ and whose influence on particles is represented by the force equation (\ref{f mu=eF mu nu v_nu}) is called the 
{\it electromagnetic} field.
The tensor $F_{\mu\nu}$ is referred to as the electromagnetic {field tensor}, or 
{field strength}.
The scalar quantity $q$ will be provisionally called the {\it electric charge-coupling}.

The identity of a given classical particle is 
preserved in time.
We  assume that there {are} such particles that their coupling with the 
electromagnetic field $q$ does not vary with time,
\begin{equation}
\dot q=0\,\,.
\label
{dot e =0}
\end{equation}

In a particular frame of reference, two three-dimensional vectors, ${\bf E}$ and ${\bf B}$, can be always defined in terms 
of six components of the antisymmetric tensor $F_{\mu\nu}$, 
\begin{equation}
E_i= F_{0i}=F^{i0}\,\,,
\label
{E_i= F_0i}
\end{equation}
\begin{equation}
B_k=-{\frac12}\,\epsilon_{klm} F^{lm}\,\,,
\label
{B_k=-frac12epsilon_klm F lm}
\end{equation}
where $\epsilon_{klm}$ is the three-dimensional Levi-Civita symbol.
Combining (\ref{E_i= F_0i}) and (\ref{B_k=-frac12epsilon_klm F lm}) with the relations 
${f^\mu}=\gamma\,({\bf F}\cdot{\bf v}\,,{\bf F})$ and $v_\mu=\gamma\,(1\,,-\,{\bf v})$, we return  to Eq.~(\ref{lorentz tree-force}).
Both quantities  $f^\mu$ and \textbf{F}, expressed 
respectively by (\ref{f mu=eF mu nu v_nu}) and (\ref{lorentz tree-force}),
go under the general name of {Lorentz force}. 

We are thus led to the unambiguous theoretical concept of electromagnetic field.
The adoption of this concept is tantamount to stating that the {\it simplest} analytic form of the four-force $f^\mu$, {\it linear} in the four-velocity $v^\mu$ and proportional to a {\it scalar} coupling $q$, is indeed feasible in our world, 
whence it follows the existence of  
a field whose states are specified by an antisymmetric tensor $F_{\mu\nu}$ in every point of spacetime.

The reader may wonder whether the simplicity is a serious theoretical argument. 
To answer this question, we recall that many fundamental physical principles are 
statements about extreme values of some physical quantities, as exemplified by the principle of {least action} in mechanics or the second law of thermodynamics selecting the state of maximal entropy.
The simplicity is a kind of extremal qualities, and so the search for the {simplest} form of  $f^\mu$ is fully justified.

On the other hand, if we would look for the simplest form of the three-force  \textbf{F}   defined in a particular frame of reference by the conventional decomposition ${f^\mu}=\gamma\,({\bf F}\cdot{\bf v}\,,{\bf F})$, then this attempt would foiled.
Indeed, \textbf{F} need not be velocity-dependent, so that the simplest form of this quantity is 
${\bf F}=$ const. 
Next in order of simplicity to constant is \textbf{F}
which only depends on space: ${\bf F}={\bf F}({\bf x})$.
We thus see that Eq.~(\ref{lorentz tree-force}) 
defies all attempts of elevating it to a  satisfactory definition of the electromagnetic field. 
The concept of electromagnetic field,  naturally arising in the Minkowski paradigm of spacetime, seems to be devoid of theoretical motivations beyond this context. 

Consider a more complex case that the interaction of a particle and electromagnetic field is specified by a {\it pseudoscalar} coupling 
$q^\star$.
Let us define the field ${}^\ast\! F^{\mu\nu}$ 
{dual} to $F^{\mu\nu}$ as
\begin{equation}
{}^\ast\! F^{\mu\nu}=\frac12\,\epsilon^{\mu\nu\rho\sigma}F_{\rho\sigma}\,\,,
\label
{ast F mu nu def}
\end{equation}
where $\epsilon^{\mu\nu\rho\sigma}$ is the four-dimensional Levi-Civita symbol, and adopt the convention that 
the three- and four-dimensional Levi-Civita symbols are related by 
\begin{equation}
\epsilon_{ijk}= \epsilon^{0ijk}\,\,.
\label
{epsilon ijk=epsilon 0ijk}
\end{equation}

The four-force $f^\mu$ {linear} in $v^\mu$, which
depends on ${}^\ast\! F^{\mu\nu}$, rather than $ F^{\mu\nu}$, is 
\begin{equation}
{f^\mu}= q^\star v_\nu {}^\ast\! F^{\mu\nu}\,\,.
\label
{f mu=e star F mu nu v_nu}
\end{equation}
Using  (\ref{E_i= F_0i}), (\ref{B_k=-frac12epsilon_klm F lm}), 
(\ref{ast F mu nu def}), and (\ref{epsilon ijk=epsilon 0ijk}), we obtain from 
(\ref{f mu=e star F mu nu v_nu}) 
\begin{equation}
{\bf F}= q^\star\left({\bf B}-{\bf v}\times{\bf E}\right). 
\label
{F =e star B - v times E}
\end{equation}

Comparison of (\ref{lorentz tree-force}) and (\ref{F =e star B - v times E}) shows
that the parameter $q^\star$ transforms as a  pseudoscalar under space reflections  
${\bf x}\to -{\bf x}$.
The quantity $q^\star$ is called the {\it magnetic 
charge-coupling}.
Particles that are affected by {\bf F} of the form
(\ref{F =e star B - v times E}) are referred to as {magnetic monopoles}.
An important point is that equation (\ref{f mu=e star F mu nu v_nu}) is not designed for giving an alternative definition of the electromagnetic field; 
this equation is to define a particle carrying 
the magnetic charge-coupling $q^\star$.
Despite prodigious experimental efforts  that went into searching for
magnetic monopoles, no manifestation of particles with 
magnetic charges $q^\star$ is found.
Presently such particles have the status of hypothetical objects.

Note, however, that the existence of just one magnetically charged particle would provide a plausible explanation for the quantization of electric charge \cite{Dirac}.
Up to now, there exists no other convincing explanation of why there is a minimal value of 
$q$, the {\it elementary} charge, and every charged  particle carries a charge which is a multiple of this elementary charge.

\section{Maxwell's equations}
\label
{Maxwell}
In the preceding section we defined the electromagnetic field as a physical object that manifests itself through its influence on a particle by the four-force linear in the particle four-velocity, provided that the coupling between this object and the affected particle is given by a scalar parameter $q$. 
This definition implies that the {state} of electromagnetic field at each spacetime point 
$x^\mu=(t,{\bf x})$ is characterized by an antisymmetric tensor $F_{\mu\nu}$.
In a particular Lorentz frame, this is equivalent to assigning the electric field intensity ${\bf E}$ and the magnetic induction ${\bf B}$ to each point.
Now let us discuss the law governing the electromagnetic field behavior in space and time.
It is well known that this law can be written as a system of partial differential equations, 
{Maxwell's equations}.
Our task is to understood the geometrical and physical contents of Maxwell's equations.
To achieve this aim it would be reasonable to attempt to derive these equations from the geometrical properties of our world, in particular from the fact that {space has three dimensions}.

Let us fix a particular frame of reference
and consider the spatial behavior of  ${\bf E}$ and ${\bf B}$. 
A smooth vector function  ${\bf V}({\bf x})$ can be reconstructed with the 
knowledge of 9 components of its gradients  $\nabla_i V_j$.
However, to do this requires actually much less information when taken into account the famous Helmholtz theorem: If a smooth vector function ${\bf V}({\bf x})$ 
disappears at infinity, it can be reconstructed from its curl 
${\nabla}\times{\bf V}$ and divergence ${\nabla}\cdot{\bf V}$\footnote{The proof of this theorem is simple.
The relation
$
{\nabla}\times({\nabla}\times{\bf V})={\nabla}\,({\nabla}\cdot{\bf V})-
\nabla^2\,{\bf V},
$
familiar from any course of the vector analysis, can be rewritten as the 
{Poisson equation}
$
\nabla^2\,{\bf V}={\bf S}
$                                          
with a computable function in the right side ${\bf S}={\nabla}\,D-
{\nabla}\times{\bf C}$.
The general solution to the Poisson equation is 
${\bf V}={\bf V}_\ast+{\bf V}_0$ where ${\bf V}_\ast$ is a particular solution of this equation, and ${\bf V}_0$ the general solution of the associated homogeneous equation $\nabla^2\,{\bf V}=0$, the Laplace equation. 
If we require that ${\bf V}$  disappear at infinity, then ${\bf V}_0$ is identically zero because, in any spatial
domain,  ${\bf V}_0$ takes its maximal and minimal values on the boundary of this domain,
in particular, given a  solution of the Laplace equation ${\bf V}_0$ vanishing on the boundary, it is identically 
zero in this domain.
Therefore, the Poisson equation has
the unique solution ${\bf V}({\bf r})={\bf V}_\ast({\bf r})=\int d^3x\,\frac{{\bf S}({\bf x})}{|{\bf r}-{\bf x}|}$ which goes to zero as $r\to\infty$
This completes the proof. }.
If we further assume that the evolutionary law for the electromagnetic field is
given by a system of differential equations, then, in view of the Helmholtz theorem, these equations must contain only those linear combinations of spacetime derivatives $\partial_\lambda F_{\mu\nu}$ which are expressed in terms of curls and
divergences of ${\bf E}$ and ${\bf B}$.
With some simple but lengthy algebra, taking into account the definitions (\ref{E_i= F_0i}), (\ref{B_k=-frac12epsilon_klm F lm}), (\ref{ast F mu nu def}), and the convention (\ref{epsilon ijk=epsilon 0ijk}), we obtain the desired linear combinations  
\begin{equation}
\partial_\mu F^{\mu\nu}=\left({\nabla}\cdot{\bf E},-{\dot{\bf E}}
+{\nabla}\times{\bf B}\right)
\label
{Maxw-vect-sec}
\end{equation}                                          
and
\begin{equation}
\partial_\mu{}^\ast\!F^{\mu\nu}=
\left({\nabla}\cdot{\bf B},\,-{\dot{\bf B}}-{\nabla}\times{\bf E}\right),
\label
{Maxw-vect-first} 
\end{equation}                                          
where the overdot denotes the time-derivative.
It is easy to verify that there are no linear combinations of $\partial_\lambda F_{\mu\nu}$ containing curls and
divergences of ${\bf E}$ and ${\bf B}$ other than those shown in (\ref{Maxw-vect-sec}) and (\ref{Maxw-vect-first}).

Thus, from essentially geometric considerations we come to the field equations
\begin{equation}
\partial_\lambda F^{\lambda\mu}=j^\mu\,\,,
\label
{Maxwell-geomt-2}
\end{equation}
\begin{equation}
\partial_\mu{}^\ast\!F^{\mu\nu}=m^\mu\,\,,
\label
{Maxwell-geomt-1}
\end{equation}
where $m^\mu$ and $j^\mu$ stand for the sources of local variations of the 
field state.
However, the meaning and construction of $m^\mu$ and $j^\mu$ are so far indeterminate.

We then adopt three additional postulates of non-geometric nature which lead directly and unambiguously to Maxwell's
equations:
\begin{quote}
\noindent
(i)  {\it Linearity  of the evolutionary law for the electromagnetic field};

\noindent
(ii)  {\it Extended action--reaction principle}; 

\noindent
(iii) {\it Lack of magnetic monopoles}.
\end{quote}
From (i), together with Poincar\'e invariance, we conclude that the sources
${j}^\mu$ and  $m^\mu$ in the field equations (\ref{Maxwell-geomt-2}) {\it and} (\ref{Maxwell-geomt-1}) are independent of $F^{\mu\nu}$. 
They may depend on particle characteristics,
such as coupling constants  $q$ and world line variables  $z^\mu({s})$. 
To find this dependence, we apply  $\partial_\mu$ to (\ref{Maxwell-geomt-2}) 
and take into account the antisymmetry of  $F^{\lambda\mu}$ and
 symmetry of  
$\partial_\mu\partial_\lambda$ in ${\lambda}$ and ${\mu}$, 
which yields the identity $\partial_\mu\partial_\lambda F^{\lambda\mu}=0$.
Therefore, the
consistency of (\ref{Maxwell-geomt-2}) is ensured by
\begin{equation}
\partial_\mu j^\mu=0\,\,.
\label
{current-conserv}
\end{equation}                                          
Integrating this equation over a domain bounded by two spacelike hypersurfaces 
$\Sigma_1$ and $\Sigma_2$ and using the Gauss--Ostrogradski{\u\i} theorem,
one can verify that the quantity
\begin{equation}
Q=\int_{\Sigma}d\sigma_\mu\,j^\mu\,\,,
\label
{charge-tot-def}
\end{equation}                                          
called the {\it total electric charge-source}, is independent of the hypersurface $\Sigma$. 
In particular, if the hypersurface  $\Sigma$ is shifted along the time axis, $Q$ remains invariant.
The relation $Q={\rm const}$, expressing the  conservation of the total
charge-source, is tempting to relate to the constancy of the charge-coupling 
$q$, implied by equation (\ref{dot e =0}).
To do this requires postulate (ii).
Indeed, let the hypersurface  $\Sigma$ be intersected by $N$ world lines of 
charged particles.
Then, by (ii),  
\begin{equation}
Q=\sum_{I=1}^N\,q_I\,\,.
\label
{charge-source=charge-coupling}
\end{equation}                                          
Imagine for a while that only a single point particle with the coupling $q$ is in the universe, then
\begin{equation}
Q=q\,\,.
\label
{charge-single}
\end{equation}                                          
Therein lies the extended action--reaction principle in electrodynamics: the charge-coupling measures the variation of the particle state for a given electromagnetic field state while the charge-source measures the variation of the electromagnetic field state for a given particle state. 
The four-dimensionality of spacetime is favorable to the action--reaction principle  because the number of degrees of freedom of electromagnetic field at each spacetime point, given by 6 components of $F_{\mu\nu}$, equals the phase space dimension of an interacting particle spanned by  3 coordinates of its position ${\bf x}$ and 3 components of conjugate momentum ${\bf p}$.
For comparison, in a conceivable $(1+5)$-dimensional pseudo-Euclidean spacetime in which the defining equation (\ref{f mu=eF mu nu v_nu}) is taken to be valid, 
the number of degrees of freedom of electromagnetic field $F_{\mu\nu}$ is $\frac12\,{d(d-1)}=15$ while the conventional phase space has dimension $2(d-1)=10$, and hence the action--reaction principle fails\footnote{The mathematically inclined reader will recognize that if we require that the electromagnetic sector of the action is  preserved in the original  Maxwellian form for every spacetime dimension $d$, then in order that the theory as a whole to be consistent, the particle sector of the action must be supplemented by terms with higher derivatives.
For example, in the above $d=6$ realm, acceleration-dependent terms are required.
Then the extended phase space, equipped with 5 constraints, has dimension $4(d-1)-5=15$, and the number of components of $F_{\mu\nu}$ equals the number of mechanical degrees of freedom, so that the action--reaction principle holds. 
The necessity of amending the particle dynamics by the addition of higher derivative terms to the action to obtain a consistent theory in the case  $d>4$ was also argued in 
Ref.~\cite{k99}, which, however, used another line of reasoning.}. 

Both quantities, $Q$ and $q$, would be reasonable to lump together as the 
{\it electric charge} or briefly the {\it charge}.

How could we realize  (\ref{charge-source=charge-coupling}) mathematically?
We assume that the charge is an inherent 
characteristic of point particles.
Then the source $j^\mu$ (called the {four-current density of electric 
charges} or simply the  
{four-current}), is given by
\begin{equation}
j^{\mu}(x)=\sum_{I=1}^N q_I\int_{-\infty}^\infty\! ds_I\,v_I^\mu(s_I)\,
\delta^{(4)}\bigl[x-z_I(s_I)\bigr]\,\,, 
\label
{j-mu-def}
\end{equation}           
where $v_I^\mu(s_I)$ is the four-velocity of $I$th particle, 
and $\delta^{(4)}\left(x\right)$ is the four-dimensional Dirac 
delta-function. 
Because the hypersurface $\Sigma$ in (\ref{charge-tot-def}) is 
arbitrary, we take $\Sigma$ such that all the world lines are perpendicular to 
it at intersection points. 
For a small
vicinity of the intersection point, we have  $ds_I\,d\sigma_\mu v_I^\mu=
d^4x$,  where $x^\mu$  are
coordinates in the Lorentz frame with the time axis directed along
 $v_I^\mu$. 
Inserting  (\ref{j-mu-def})  in (\ref{charge-tot-def}), we arrive at
(\ref{charge-source=charge-coupling}).   

We next turn to equation  (\ref{Maxwell-geomt-1}) and reiterate mutatis mutandis the above arguments.
By the action-reaction principle, the total magnetic charge-source 
$Q^\star=\int d\sigma_\mu\, m^\mu$
equals the sum of magnetic charge-couplings,
$Q^\star=\sum_{I=1}^N q^\star_I$.
With 
postulate (iii), we find  
\begin{equation}
m^\mu=0\,\,.
\label
{lack-monopole}
\end{equation}                                          

Finally, the electromagnetic field is governed by the system of equations
\begin{equation}
\partial_\lambda F^{\lambda\mu}=4\pi j^\mu\,\,,
\label
{Maxw-second}
\end{equation}                                          
\begin{equation}
\partial_\lambda{}^\ast\!F^{\lambda\mu}=0\,\,,
\label
{Maxw-first}
\end{equation}                                          
which are just Maxwell's equations.

To summarize, a major part of information encoded in Maxwell's equations (\ref{Maxw-second}) and (\ref{Maxw-first}) is taken from global topological properties of spacetime, notably from the fact that our world has dimension $d=4$, and the residual information, seemingly divorced from geometry, which represents the physical contents of equations (\ref{Maxw-second}) and (\ref{Maxw-first}) translates into the above assumptions (i)--(iii).

\section{Conclusion and outlook (for the expert reader)}
\label
{concluding}
The idea of {spacetime} provides a  satisfactory framework for the unique definition of the electromagnetic field.
This definition rests on the theoretical appeal of the {\it simplest} form of the four-force $f^\mu$ and the phenomenological fact that this simplest option is indeed taken by nature, whence it follows that there exists a  physical object whose states are specified by an antisymmetric tensor $F_{\mu\nu}$ in every point of spacetime.
It is this object which is regarded as the  electromagnetic field.
It is remarkable that the same definition is applicable to a world with  space dimension other than 3.

One can extend the consideration of the four-force $f^\mu$ linear in the four-velocity $v^\mu$ 
to the case that the coupling  $Q^a$ is a {\it vector} in some ${\cal N}$-dimensional vector space describing charge states of the  particle, to yield the force equation
\cite{Wong} 
\begin{equation}
{f^\mu}
=\sum_{a=1}^{\cal N}\,Q^a v_\nu G_a^{\mu\nu}\,\,,
\label
{f mu=YMW}
\end{equation}
where the quantity $G^a_{\mu\nu}$ is called the {\it Yang--Mills field} strength.
A refined version of this vector coupling, involving the Lie algebra structures, gains insight into the form of gauge
fields mediating the weak and strong interactions. 

If we want the gravitational interaction to be covered in this  framework, we should turn to the four-force $f^\mu$ which is  {\it quadratic} in the four-velocity $v^\mu$.
The necessity of considering quadratic relationships is due to the empirical fact that identical gravitating particles attract each other. 
With reference to Exercise 7.2 in \cite{Misner}, it transpires that if the interaction is carried by a field  $F_{\mu\nu}$ (or $G^a_{\mu\nu}$), then
identical particles with a real coupling $q$ (or $Q^a$) repel each other.
In contrast, {quadratic} relationships between 
$f^\mu$ and $v^\mu$ are compatible with 
scalar and symmetric tensor fields mediating such interactions which ensure that identical particles are attracted together (Exercises 7.1 and 7.3 in \cite{Misner}).
A diligent student may wish to explore the general quadratic relationship 
\begin{equation}
f_\lambda=-\Gamma_{\lambda\mu\nu}v^\mu v^\nu\,\,,
\label
{quadr-v-force}
\end{equation}
where $\Gamma_{\lambda\mu\nu}$ is an arbitrary rank $(0,3)$ tensor, and show that $\Gamma_{\lambda\mu\nu}$ is always identically vanishing, except for the case that spacetime is a curved  pseudo-Riemannian manifold with the line element $ds^2$ given by
\begin{equation}
ds^2=g_{\mu\nu}(x)dx^\mu dx^\nu\,\,,
\label
{Riemann-metric}
\end{equation}
where $g_{\mu\nu}$ is a symmetric rank $(0,2)$ tensor interpreted as the metric of this manifold, and 
$\Gamma_{\lambda\mu\nu}$ is expressed in terms of $g_{\mu\nu}$ as
\begin{equation}
\Gamma_{\lambda\mu\nu}=\frac12\left(\partial_\mu  g_{\nu\lambda}+\partial_\nu  
g_{\lambda\mu}-\partial_\lambda g_{\mu\nu}\right).
\label
{christoffel-lower}
\end{equation}
Therefore, in an effort to describe the gravitational interaction, we are inevitably led to the idea of spacetime warping.

The derivation of the law governing  
the electromagnetic field from `next to nothing' (or, more precisely, from the properties of  four-dimensional spacetime) can be an enormously enlightening experience in the physics curriculum.
When it is considered (\ref{Maxw-vect-sec}) and (\ref{Maxw-vect-first}), and the four-current 
is decomposed as  $j^\mu=(\varrho,\,{\bf j})$, one recasts Maxwell's equations (\ref{Maxw-second}) and (\ref{Maxw-first}) in the conventional three-dimensional vector form 
\begin{equation}
{\nabla}\cdot{\bf E}=4\pi\varrho\,\,,
\label
{Maxw-div-E}
\end{equation}                                          
\begin{equation}
{\nabla}\times{\bf B}=4\pi{\bf j}+{\dot {\bf E}}\,\,,
\label
{Maxw-rot-B}
\end{equation}                                          
\begin{equation}
{\nabla}\cdot{\bf B}=0\,\,,
\label
{Maxw-div-B}
\end{equation}                                          
\begin{equation}
{\nabla}\times{\bf E}=-{\dot {\bf B}}\,\,. 
\label
{Maxw-rot-E}
\end{equation}                                          

One could hardly discern the geometric origin of these equations.
This brings up the question of whether there is   space dimension other than 3 for which the Lorentz force  equation (\ref{lorentz tree-force}) and Maxwell's equations (\ref{Maxw-div-E})--(\ref{Maxw-rot-E}) can be written in a similar vector form.
Apart from ${\mathbb{E}_3}$, the {\it cross product} of two vectors can only be defined  in  
six-dimensional Euclidean space ${\mathbb{E}_7}$.
It is well known that the cross product in 
${\mathbb{E}_{3}}$ is closely related to quaternion algebra.
The opportunity to define the cross product in 
${\mathbb{E}_{7}}$ arises from octonion  algebra, the largest composition algebra.
Although this opportunity leads to the  reproduction of  Maxwell's equations (\ref{Maxw-div-E})--(\ref{Maxw-rot-E}), the Lorentz force equation (\ref{lorentz tree-force}) is modified by the addition of an extra term \cite{Silagadze}, so that the geometric and physical contents of the resulting equations are quite different from what was just represented.
   
Separating the geometric content of Maxwell's equations from their physical content allows to appreciate features peculiar to electrodynamics 
in comparison with the three other fundamental interactions.
Take, for example,  the action--reaction principle.
In general relativity this principle does not hold. 
Indeed, from the equation of motion for a point particle 
\begin{equation}
\frac{d^2z_\lambda}{d\tau^2}+\Gamma_{\lambda\mu\nu}\frac{dz^\mu}{d\tau}\frac{dz^\nu}{d\tau}=0\,\,,
\label
{geodesics}
\end{equation}
where $\Gamma_{\lambda\mu\nu}$ is the Christoffel symbols assigned to the curved manifold in which the particle is located, and  $\tau$ an appropriate evolution variable, 
we see that the gravitational field exerts on all particles in a
uniform way no matter what their masses.
On the other hand, consider the  equation of gravitational field  
\begin{equation}
R^{\mu\nu}-\frac12Rg^{\mu\nu}=8\pi G_{\rm N} T^{\mu\nu}\,\,,
\label
{Hilbert-Einstein} 
\end{equation}
in which the source of gravitational field is taken to be the stress-energy tensor of a
single particle with gravitational mass $m_{\rm g}$ moving along the world line
$z^\mu(\tau)$, 
\begin{equation}
T^{\mu\nu}(x)=m_{\rm g}\int_{-\infty}^\infty d\tau\,{\dot z}^\mu(\tau)\,{\dot z}^\nu(\tau)\,
\delta^{(4)}[x-z(\tau)]\,\,.
\label
{ideal gas}
\end{equation}
The greater is $m_{\rm g}$, the  stronger is the spacetime warping.
Thus, the influence of particles with different gravitational masses on the state of gravitational field is different.  
This is  contrary to the action--reaction principle.

Maxwell's equations in covariant tensor form  (\ref{Maxw-second}) and (\ref{Maxw-first}) 
offer a universal description of the  behavior of 
electromagnetic field  in a world  with an
arbitrary  space dimension $d-1={n}$.
At first glance this is  a disadvantage, rather than advantage, 
because this writing of Maxwell's equations does not suggest that the ${n}=3$ stands out against
other values of ${n}$, and hence, the above argument regarding the three-dimensional 
origin of Maxwell's equations seems fail.
But this is not the case.
The `${n}=3$ birthmark' is implicit in (\ref{Maxw-second}) and (\ref{Maxw-first}): 
these equations are invariant under the group of {\it conformal} transformations
(the largest group of spacetime symmetry of electrodynamics \cite{Bateman}, 
\cite{Cunningham}) {\it only} in the case ${n}=3$ \cite{Weyl}.

The separation of the geometric and non-geometric constituents of electrodynamics makes it  possible to test each constituent separately by looking into its modifications.
For example, if we abandon the linearity,  then we come to non-linear 
modifications of  electrodynamics  such as the Born--Infeld theory \cite{Born-Infeld}.  
However,  among those non-linear modifications of  electrodynamics with a reasonable weak field limit yielding the Maxwell theory, there is no such to exhibit conformal invariance of the field equations.
This fact appears to be the strongest argument in support of the linearity.

As to assumption (iii), reflecting the  evidence that particles with magnetic charges $q^\star$ 
do not exist (or perhaps are extremely rare in nature), it is also related to geometry but in a subtle way.
To see this, we refer to the fact that the homogeneous equation (\ref{Maxw-first}) can be derived from the symplectic structure of Poisson manifolds
\cite{Dyson}, \cite{Carinena}.

To summarize, the physical content of Maxwell's equations can be reformulated in geometrical terms.
However, this reformulation would tell us about  structures which are not identical to those inherent in the four-dimensional geometry of our spacetime.
In fact, these two geometries are superimposed in electrodynamics.

\end{document}